\documentclass[preprint,twocolumn,tightenlines,aps,prb,10pt,floats,floatfix]{revtex4}
\usepackage{psfig}
\usepackage{epsfig}
\usepackage{graphicx}
\usepackage{dcolumn}
\usepackage{longtable}

\def\erf{\mbox{\rm{erf}}}

\begin{document}

% \title{Langevin dynamics with space-dependent friction arising from 
% a secondary stochastic force}
\title{Dissipating the Langevin equation in the 
presence of an external stochastic potential}
\author{Jeremy M. Moix}
\author{Rigoberto Hernandez}
\thanks{Author to whom correspondence should be addressed} 
\email{hernandez@chemistry.gatech.edu.}
\affiliation{Center for Computational and Molecular Science and Technology, \\
School of Chemistry and Biochemistry, \\
Georgia Institute of Technology, \\
Atlanta, GA  30332-0400}
\date{\today}
\begin{abstract}%
In the Langevin formalism, 
the delicate balance maintained between the fluctuations
in the system and their corresponding dissipation may be 
upset by the presence of a secondary, space-dependent stochastic force, 
particularly in the low friction regime. 
In prior work, the latter was dissipated self-consistently
through an additional uniform (mean-field) friction
[Shepherd and Hernandez, {\it J.~Chem.~Phys.},
{\bf 115}, 2430-2438 (2001).]
An alternative approach to ensure that equipartition is satisfied 
relies on the use of a space-dependent friction while ignoring nonlocal
correlations.
The approach is evaluated with respect to its ability to
maintain constant temperature for two simple one-dimensional, 
stochastic potentials of mean force wherein the friction can
be evaluated explicitly when there is no memory in the barriers.
%in the memory-less limit.
The use of a space-dependent friction is capable of 
providing qualitatively similar results to those obtained previously, 
but in extreme cases, deviations from equipartition may be observed
due to the neglect of the memory effects present in the stochastic potentials.
\end{abstract}

\maketitle

\section{Introduction}

In the theory of diffusion processes over fixed barriers, 
numerous studies have shown that the dissipative
term in the Langevin equation is rarely constant along the reaction 
coordinate.\cite{carm83,sbb88,zhu88,krishnan92,bere93,straus93b,neria96,schwartz99}
%\marginpar{citations?}
A general rate theory when the friction is both
space- and time-dependent has been developed to account for this
phenomenon over the entire friction regime.\cite{haynes94,haynes93,voth92b}
One might na\"{i}vely expect that a space-dependent 
component must be included in the friction kernel to capture the essential
dynamics of a given system.  
However, this is not always the case.  
Several groups have shown that the average dynamical properties
may still be adequately described by a 
generalized Langevin equation with space-independent friction
even when the reaction 
coordinate has a strong spatial dependence.\cite{straus93b,sbb88,hynes87,krishnan92}
%\marginpar{CITATIONS?}
An analysis by Haynes and Voth 
concluded that the key factor is not whether
the friction is space-dependent, since it generally will be, 
but rather how the friction varies 
along the reaction coordinate.\cite{voth95}
In particular, they suggest that the symmetry of the space-dependent friction
with respect to the barrier can be used as a metric for evaluating the 
role of the friction in the dynamics. 
Similar product and reactant states will give 
rise to similar (symmetric) friction
components about the transition state.
Perhaps surprisingly, 
an antisymmetric friction does not have a significant impact on the 
dynamics,
while a symmetric friction can result in large deviations from the 
predictions of standard rate theories for processes with space-independent 
friction.\cite{voth95,straus93b,zhu88,sbb87,krishnan92,schwartz99}
Thus, the Langevin model with a uniform effective friction can often 
approximate the dynamics of projected variables even if the formal 
projection would have required a space-dependent model.

The central question explored in this work is whether a single
uniform effective friction suffices even when the Langevin system is
subjected to an external space-dependent stochastic potential.
%
%In the present work, 
%the main concern is the description of the space-dependence in the 
%friction that may arise from fluctuations in an external
%(but space-dependent) stochastic potential that is symmetric
%about a given barrier,
%rather than from the space-dependent coupling to the thermal bath.
%With periodic potentials of mean force, 
%the friction is perfectly symmetric with respect to a given barrier, 
%and may give rise to interesting dynamical effects according to
%previous studies.
%
The behavior of a Brownian particle diffusing across 
various subsets of this class of potentials has been
the subject of intense research.\cite{DG92,march98,pollak93a,
pollak99,astumian98,VDB93,Reim95,Hang95b,lind94}
This activity has 
largely been motivated by the discovery of resonant activation
in which the rate of transport over a stochastic barrier
exhibits a maximum as a function of the correlation time in
the fluctuations of the barrier height.\cite{DG92}
However,
until recently, simulations of these systems have not been performed in
the low friction regime, where deviations from
equipartition may occur,
due to an inability to adequately
describe the friction in the presence of an additional
stochastic force.\cite{hern00c,hern02c}
In previous work, 
the dissipation of this excess energy 
was achieved through a self-consistent approach in which
the friction constant is renormalized iteratively 
until equipartition is satisfied.\cite{hern00c}
This renormalization is approximate because it does not explicitly
account for the correlations between the external stochastic forces 
across space and time, but rather uses a single mean friction
to dissipate theses forces at times longer than their correlation times.
%
% However, 
% an analytic form for this contribution to the friction 
% has not yet been found. 
% Here, 
A possible improvement to the self-consistent approach can be obtained 
by allowing the friction to be space-dependent while 
explicitly ignoring the memory in the stochastic potential,
In the special case that the stochastic potential has no memory, then
this treatment is exact.
However, this approximation is often not justified when modeling real 
systems and therefore, the model potentials employed are chosen to have an
exponentially decaying memory of their past states.  
In the most extreme cases, these correlations can result in deviations from 
equipartition during the course of the simulation, although the space-dependent 
friction dissipates such fluctuations correctly in most situations.
The general conclusion appears to be that the more detailed space-dependent
approach is in qualitative agreement with the self-consistent approach
and hence, as in the fixed barrier case, Langevin systems with 
stochastic forces may be dissipated by a single (though renormalized)
uniform friction.

The conclusions of this work are supported by a study of two different
classes of one-dimensional problems in which the particle diffuses
across a periodic array of coherent or incoherent barriers.
These two cases can be specified by sinusoidal or 
merged-harmonic-oscillator potentials, respectively.
For such simple forms of the stochastic potential, analytic expressions
for the friction as a function of the spatial coordinate can 
readily be obtained and are presented in Sec.~\ref{sec:formalism}.
The resulting Langevin dynamics across these potentials 
dissipated either uniformly or through the space-dependent
friction are illustrated in Sec.~\ref{sec:results}.

\section{Langevin Model with Stochastic Potentials}\label{sec:formalism}

An equation of motion describing the diffusion of a particle  
influenced by a stochastic potential of
mean force can be adequately described by a
phenomenological Langevin equation of the form,
\begin{equation}
\dot{v}=-\gamma(t) v
        +\xi(t)
        +F(x;t)
        \;, \label{eq:eqm}
\end{equation}
where $F(x;t) \equiv -\nabla_{x}U(x;t)$ is an external stochastic force,
and $\gamma(t)$ is the friction required to dissipate
both the thermal forces and those due to the external stochastic potential.
The thermal bath is described by $\xi(t)$, 
which is a Gaussian white noise source
with time correlation given by the 
fluctuation-dissipation relation (FDR),
\begin{equation}
\langle\xi(t)\xi(t^{\prime})\rangle=2k_{\rm B}T\gamma_{\rm th}
\delta(t-t^{\prime})
\;.\label{eq:xi_xi}
\end{equation} 
In the limit that $F(x;t)=F(x;0)$ for all $t$, %(what if F(x;t)=F(x))
these equations reduce to the 
Langevin equation with $\gamma(t)=\gamma_{\rm th}$.
Otherwise, the question remains as to what is the appropriate form
of $\gamma(t)$.
Two approaches for addressing this question are presented 
in Sections~\ref{sec:uniform} and~\ref{sec:sdf},
after first describing the explicit forms of the stochastic
potentials.

\subsection{Stochastic Potential Representation}\label{sec:pot}

The space-dependent friction (SDF) that arises from the 
fluctuations in $F(x;t)$ can readily be evaluated analytically for
two different classes of one-dimensional
stochastic potentials.
The first of these is a sinusoidal potential taking the 
general form,
\begin{equation}
U(x;t)=\left(E_{\rm b}+\frac{1}{2}\eta(t)\right)
       \left(\sin\left(\frac{\pi x}{2}\right)+1\right)
\;,\label{eq:sin}
\end{equation}
in which the barriers fluctuate {\it coherently} with each other.
The second is constructed using a series of merged harmonic oscillators (MHOs) 
in which each barrier is allowed to fluctuate independently
({\it incoherently}) of one another, and is specified by
\begin{equation}
U(x,t)= \left\{\begin{array}{ll}
        \frac{1}{2}k_0(x-x_m^0)^2   &\mbox{for $x_m^0 < x \le x_m^-$}\\
        V_{m}^{\ddagger} + 
        \frac{1}{2}k_{m}^{\ddagger}(x-x_m^{\ddagger})^2 
                                    &\mbox{for $x_m^- < x \le x_m^+$}\\
        \frac{1}{2}k_0(x-x_{m+1}^0)^2  &\mbox{for $x_m^+ < x \le x_{m+1}^0$}

      \end{array}\right.
\;,\label{eq:mho}
\end{equation}
where the $m^{th}$ well and adjacent barrier
are centered at $x_m^0=-\lambda/2+m\lambda$ and 
$x_m^\ddagger=m\lambda$, respectively. 
The connection points are chosen to ensure
continuity in the potential and its first derivative such that
$x_{m}^{\pm}=\pm{k_0\lambda}/{(2k_{0}-2k_{m}^\ddagger)}+m\lambda$. 
As opposed to the sinusoidal potential, the width of the MHO barriers 
varies stochastically in time according to the relation 
$k_{m}^{\ddagger}=-(k_{0}+\eta(m,t))$, 
which, in turn, defines the barrier height
$V_{m}^{\ddagger}=-k_{0}k_{m}^{\ddagger}\lambda^2/(8k_{0}-8k_{m}^{\ddagger})$.
The remaining parameters in the  
potentials are chosen such that the lattice spacing is 4 and the
thermal energy of the particle is $1/6$ of the average value of 
the barrier heights.
%average barrier height for each is $6 k_{\rm B} T$ 

The stochastic term,
$\eta(t)$, is defined as an Ornstein-Uhlenbeck process governed by 
the following differential equation,
\begin{equation}
\dot{\eta}(t)=-\frac{\eta(t)}{\tau_{\rm{c}}}+
              \sqrt{\frac{2\sigma^2}{\tau_{\rm{c}}}}\zeta(t)
\;,
\end{equation}
with the probability distribution,
\begin{equation}
P\left(\eta(t)\right)=\frac{1}{\sqrt{2\pi\sigma^2}}\exp\left(-\frac{\eta(t)^2}
        {2\sigma^2}\right)
\;,\label{eq:prob}
\end{equation}
and time correlation,
\begin{equation}
\langle\eta(t)\eta(t^{\prime})\rangle=\sigma^2\exp\left(-\frac
                                      {|t-t^{\prime}|}{\tau_{\rm{c}}}\right)
\;.
\end{equation}
The variance of the distribution is given by $\sigma^2$, $\tau_{\rm{c}}$ is
the correlation time, and $\zeta(t)$ is an additional
white noise source.
The distribution of barriers heights for the sinusoidal potential
is given directly by the distribution of $\eta(t)$,
but due to the nature of the expression for the barrier heights of the MHOs, 
the resulting distribution for this potential takes on a more complex
form that is sharper and slightly skewed compared with Eq.~\ref{eq:prob}.
As a result, a much smaller range of fluctuations is allowed for the MHO
than the sinusoidal potential to ensure that the distribution 
does not become significantly non-Gaussian.  
More details on the exact behavior of the
MHO barrier heights are provided in Ref.~\onlinecite{hern00c}.

\subsection{Uniform Dissipation}\label{sec:uniform}

In previous work,\cite{hern00c} 
a self-consistent procedure was developed to ensure that the evolution
of the system using Eq.~\ref{eq:eqm} remains in thermal equilibrium.
This was accomplished through an iterative
procedure in which the friction, given by the sum of the two 
contributions from the thermal bath and the stochastic potential, $\it{i.e.}$
$\gamma \equiv  \gamma_{\rm th}+\gamma_{\rm F}$, 
is renormalized according to the relation,
\begin{equation}
\gamma^{(n+1)}=\gamma^{(n)}\left(\frac{\langle v^2(t) 
                                       \rangle_n}{k_{\rm b}T}\right)
\label{eq:SCF}
\;.
\end{equation}
The friction for the next iteration is determined from the value of the 
friction at the current step scaled by the magnitude of the 
deviation from equipartition seen in the dynamics
until convergence is reached to within a desired accuracy.
The main criticism to this approach lies in the approximation 
made in developing Eq.~\ref{eq:SCF} in which the stochastic potential
is treated as a local noise source,
$\gamma_{\rm F}$, obeying a fluctuation-dissipation relation
equivalent to Eq.~\ref{eq:xi_xi}.
However, the stochastic potentials have memory and are therefore 
nonlocal in nature leading to non-vanishing cumulants at 
third and higher orders.
These effects are included, but only in an average manner, to second order
in this approach.

\subsection{Space-Dependent Dissipation}\label{sec:sdf}
 
% An alternative approach that explicitly includes the spatial 
% dependence of the friction will be described.  

An alternative approach to dissipating the external stochastic force
relies on replacing the space- and time-dependent friction, $\gamma(x,t)$, 
by a space-dependent friction, $\gamma(x(t))$, satisfying a local 
FDR.
Given that the size of the fluctuations in $F(x;t)$
depend on $x$ at a given $t$,
a Brownian particle moving quickly across the surface will experience
a series of forces whose relative magnitudes depend on the particle's velocity.
However when the the Brownian particle moves slowly, 
the particle will sample only the local fluctuations of the
stochastic potential in the vicinity of its local position $x$.
In this regime, the particle arrives at a local quasi-equilibrium  
which must necessarily satisfy the FDR locally. 
This suggests that the dissipation should not be uniform, but rather
should depend on position, and therefore indirectly on time.
It should be noted that while the mean-field approach 
described in the previous subsection
is capable of including the average of the 
correlations between the fluctuations, 
the approximation made here does not account for any of the memory effects. 
However, in the limit that there is no 
memory in the external stochastic potential,
%\tau_{\rm{c}}=0$, 
the following results are exact.

The question now arises of how to explicitly describe the friction constant in 
the presence of an additional fluctuating force resulting from the
potentials of mean force given in Eqns.~\ref{eq:sin} and ~\ref{eq:mho}.  
The friction constant must dissipate the excess energy 
that arises from the fluctuating forces through a local 
space-dependent FDR,
\begin{equation}
2k_{\rm B}T\gamma_{\rm c}(x;t)=\langle\delta F_{\rm c}(x;t)^2\rangle
\;,\label{eq:FDT}
\end{equation}
where the cumulative force is simply the sum of the thermal Gaussian noise and
the stochastic force arising from the external potential, 
$F_{\rm c}=F_{\rm th}+F_{\rm U}$. 
Assuming the respective 
fluctuations in the bath and the potential are uncorrelated, {\it i.e.} 
$\langle\delta F_{\rm th} \delta F_{\rm U} \rangle=0$,
then Eq.~(\ref{eq:FDT}), reduces to 
\begin{equation}
2 k_{\rm B}T\gamma_{\rm c}(x;t)=\langle\xi(t)\xi(t^\prime)\rangle+
                                \langle \delta F_{\rm U}(x;t)^2 \rangle
\;,
\label{eq:fric}
\end{equation} 
%where the time dependent fluctuations in 
%the potential are implicitly contained in $\eta$.  (redundant)
The thermal fluctuations 
are ohmic as given in Eq.~\ref{eq:xi_xi}, 
and the relationship for the fluctuations in the force is
$\delta F_{\rm U}(x;t) \equiv 
 F_{\rm U}(x;t)-\langle F_{\rm U}(x;t)\rangle_{\eta}$, 
where the average is taken with respect to the auxiliary stochastic
variable, $\eta$.
The average value of the force can be determined according 
to the usual integrals, 
\begin{equation}
\langle F_{\rm U}(x;t)\rangle=
                             \frac{-\int_{-\infty}^{\infty}\!d\eta\, P(\eta)
                             \nabla_x U(x;t) }
                             {\int_{-\infty}^{\infty}\!d\eta\, P(\eta)}
\;,
\end{equation}
where the fluctuations in the force are governed by the stochastic 
Ornstein-Uhlenbeck process, $\eta$, whose probability distribution is 
given by Eq.~\ref{eq:prob}. 

The remaining steps of the derivation rely
upon the specific form of the potential.
As an illustration, the SDF is evaluated explicitly below
for the simpler sinusoidal (coherent) stochastic potential.
(The results for the incoherent MHO potential can be found in the Appendix.)
The derivation begins by direct evaluation of Eq.~\ref{eq:fric}
for the specific class of potentials.  
As remarked above, the first term reproduces 
the FDR, Eq.~\ref{eq:xi_xi}, for the thermal forces.
Ignoring the correlation in the forces at different times, the second 
reduces to:
\begin{eqnarray}
\langle \delta F_{\rm U}(x;t)^2 \rangle & = &
                    \frac{\pi^2}{4}\cos^2\left(\frac{\pi x}{2};t\right)
                    \times \nonumber \\
             &   &  \int_{-\infty}^\infty\!d\eta\,
                    \left(E_{\rm b}+\frac{1}{2}\eta\right)^2P(\eta)- \nonumber\\
             &   &  \left[\frac{\pi}{2}\cos\left(\frac{\pi x}{2};t\right)
                    \right.\times \nonumber \\
             &   &  \left. \int_{-\infty}^{\infty}\!d\eta\, 
                    \left(E_{\rm b}+\frac{1}{2}\eta\right)P(\eta)\right]^2
\;.
\end{eqnarray}
The Gaussian integrals are readily evaluated to yield:
\begin{equation}
\langle \delta F_{\rm U}(x;t)^2 \rangle=\frac{\sigma^2\pi^2}{16}
                                        \cos^2(\frac{\pi x}{2};t)
\;.
\end{equation}
Upon substitution into Eq.~\ref{eq:FDT},
the explicit form of the SDF is
\begin{equation}
\gamma_{\rm c}(x;t)=\gamma_{\rm th}\delta(t-t^\prime)+
                   \frac{\sigma^2\pi^2}{32k_{\rm B}T}\cos^2(\frac{\pi x}{2};t)
\label{eq:sdfsin}
\;.
\end{equation}
This is the simplest possible form for this result,
and is due to the separability of the potential
into a sum of deterministic and linear stochastic terms.  
In fact, it is easily shown that for any separable potential 
of the form,
\begin{equation}
U(x;t)=\bar{U}(x)+\eta(t)W(x)
\;,\label{eq:separable}
\end{equation}
where $\bar{U}(x)$ is the deterministic 
component of the potential of mean force,
then the additional friction due
to the stochastic potential is given by
\begin{equation}
\langle \delta F_{\rm U}(x;t)^2 \rangle=\left(\nabla_x W(x)\right)^2
                                        \int\!d\eta\, (\eta^2-\eta) P(\eta)
\;,
\label{eq:MomentGen}
\end{equation} 
provided the distribution is normalized.
The MHO does not satisfy the condition of Eq.~\ref{eq:separable}
and hence its friction correction can not be
obtained by Eq.~\ref{eq:MomentGen}.
The form of the friction correction for the MHO consequently
contains more terms, but the requisite approximation (that the
forces are uncorrelated at different times) enters the
derivation in a conceptually equivalent way.

\subsection{Mean First-Passage Times}

The dynamics of the system were characterized by the mean first passage
time (MFPT) of a particle to escape its initial minima and establish
a quasi-equilibrium within another well.  With periodic, stochastic 
potentials, this may be accomplished by defining a region
of the phase space of the particle bounded by an 
energetic constraint.\cite{hern02b}
The MFPT
is simply the average of a sufficient number of first passage processes into
this region,
with the corresponding rate given by the inverse of the MFPT.
While the incorporation of a space-dependent friction in the algorithm for
the numerical integration of the equations of motion would seemingly result
in a dramatic increase in computational expense, 
the actual effort is comparable to the previous mean-field approach because the 
preliminary convergence procedure for the friction constant is now unnecessary.

\section{Results and Discussion}\label{sec:results}

\begin{figure}[t]
\begin{center}
\includegraphics*[width=7.5cm]{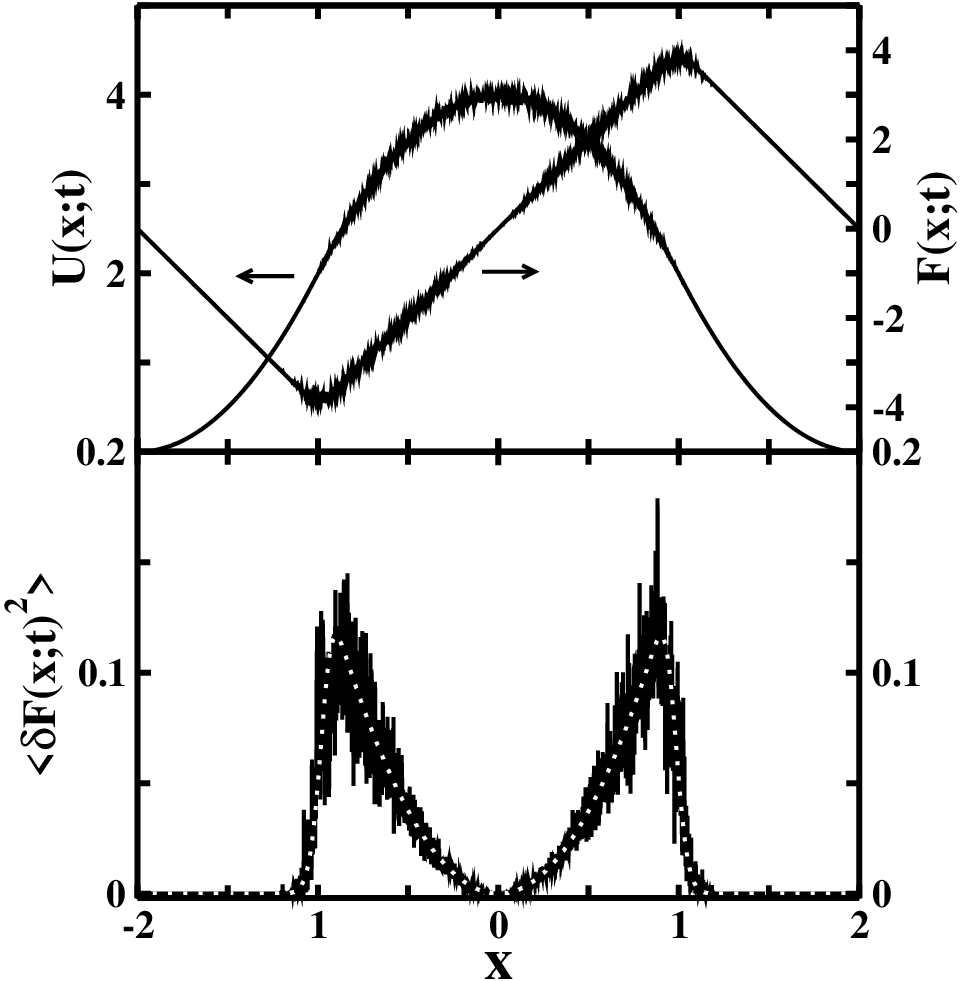}
\end{center}
\caption{
Representative fluctuations over one period of the MHO
potential and force (top panel),
and the resulting space-dependent friction (bottom panel).
The numerical component in the bottom panel
is displayed as the solid black line, with
the analytic result, given in the Appendix, as the dotted white line.
The temperature is 2/3, the variance
is 0.22, and the thermal friction is 0.08.
}
\label{fig:sdfmho}
\end{figure}

\begin{figure}[t]
\begin{center}
\includegraphics*[width=7.5cm]{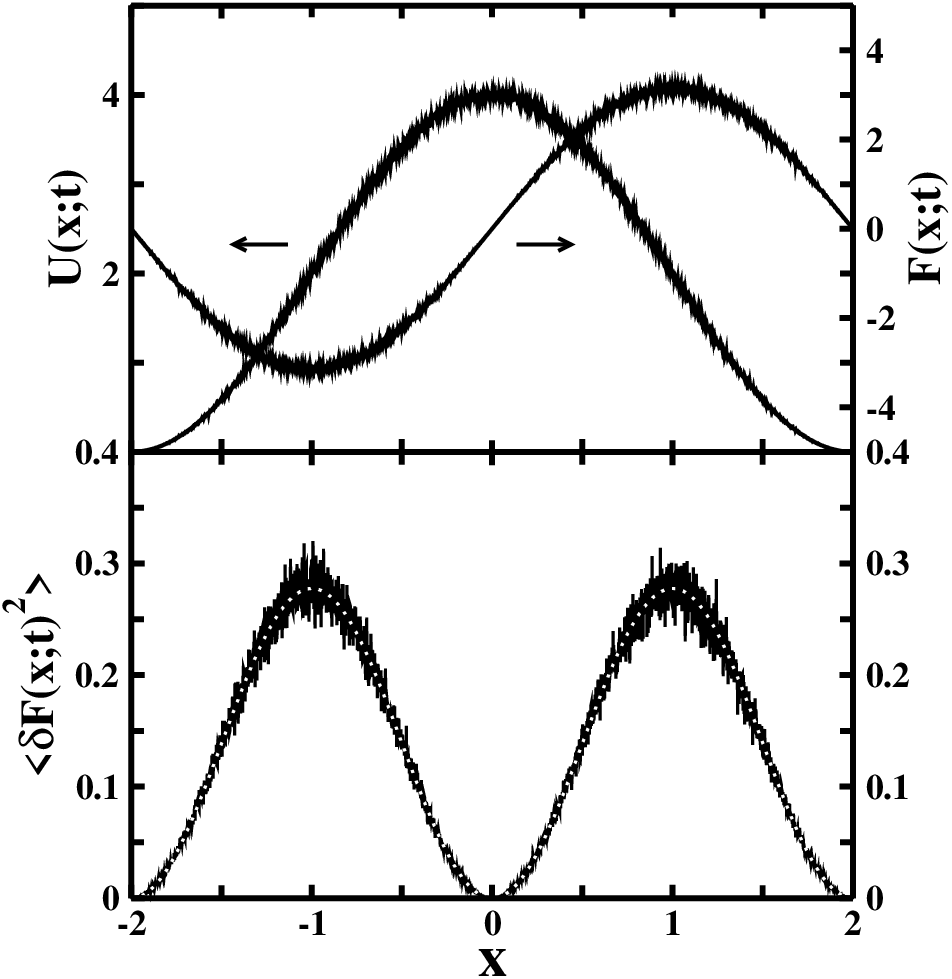}
\end{center}
\caption{
Representative fluctuations over one period of the sinusoidal
potential and force (top panel),
and the resulting space-dependent friction (bottom panel).
The numerical result is displayed as the solid black line,
with the analytic result, given by Eq.~\ref{eq:sdfsin}, shown as
the dotted white line.
The parameters used are the same as in Fig.~\ref{fig:sdfmho}.
}
\label{fig:sdfsin}
\end{figure}

The analytic and numerical space-dependent components of the friction
over one period of the MHO and sinusoidal potentials can be seen in the bottom
panel of Figs.~\ref{fig:sdfmho} and 
~\ref{fig:sdfsin}, respectively, with the numerical results
averaged over 500 representative trajectories.

%%%%%%%%%%%%%%%%%%%%%%
%\newpage
                                                                                
%%%%%%%%%%%%%%%%%%%%%%%
%%%%%%%%%%%%%%%%%%%%%%%
%/newpage
 
The top panels display the fluctuations in the potential and the resulting
forces that give rise to the space-dependent friction.
The analytic forms of the SDF, displayed as the dotted white line, 
agree with the corresponding numerical results, and exact agreement
is obtained upon further averaging.  
The fluctuations in the forces reach
a maximum at approximately the midpoint between the minima and maxima, 
where deviations from the average force take on the largest values.  
The fluctuations in the potential are largest at the barriers, 
while the forces are zero at these locations.
This leads to a vanishing contribution to the total friction 
from the space-dependent component at these points.  
In the well region, the behavior of the SDF 
for the sinusoidal and MHO potentials is inherently different.  
The SDF for the MHO is zero outside of the barrier 
region since the wells do not fluctuate by construction.
However, the sinusoidal potential fluctuates continuously throughout
leading to a friction correction along the entire reaction coordinate.  
Consequently, the magnitude of the friction correction 
in simulations employing the sinusoidal potential are
slightly larger than that in those employing the MHO.
But, as illustrated below,
this effect does not have a dramatic effect on the resulting dynamics.  

Values of the friction corrections 
calculated from the iterative and space-dependent
approaches for the MHO and sinusoidal potentials are displayed
in Table~\ref{tb:mho} with  
the values of the thermal friction listed in the left-most column.
\begin{table*}[ht]
\begin{center}
\begin{tabular}{ccccccccccccc}
\hline \\[-9truept]
\hline \\[-8truept]
\multicolumn{1}{c}{} &
\multicolumn{3}{c}{
    \underline{$
                 \mbox{ }\mbox{ }\mbox{ }\mbox{ }\mbox{ }\mbox{ }
                 \mbox{ }\mbox{ }\mbox{ }\mbox{ }\mbox{ }\mbox{ }
                 \rm{MHO}
                 \mbox{ }\mbox{ }\mbox{ }\mbox{ }\mbox{ }\mbox{ }
                 \mbox{ }\mbox{ }\mbox{ }\mbox{ }\mbox{ }\mbox{ }
               $}
}&
\multicolumn{3}{c}{
   \underline{$
                 \mbox{ }\mbox{ }\mbox{ }\mbox{ }\mbox{ }\mbox{ }
                 \mbox{ }\mbox{ }\mbox{ }\mbox{ }\mbox{ }\mbox{ }
                 \rm{Sin}
                 \mbox{ }\mbox{ }\mbox{ }\mbox{ }\mbox{ }\mbox{ }
                 \mbox{ }\mbox{ }\mbox{ }\mbox{ }\mbox{ }\mbox{ }
   $}
}\\
%\multicolumn{3}{c}{
%   \underline{$
%                 \mbox{ }\mbox{ }\mbox{ }\mbox{ }\mbox{ }\mbox{ }
%                 \mbox{ }\mbox{ }\mbox{ }\mbox{ }\mbox{ }\mbox{ }
%                 \tau_{\rm c}=10^1
%                 \mbox{ }\mbox{ }\mbox{ }\mbox{ }\mbox{ }\mbox{ }
%                 \mbox{ }\mbox{ }\mbox{ }\mbox{ }\mbox{ }\mbox{ }
%   $}
%}\\
                                                                                
$\gamma_{\rm th}$ &
                                                                                
$ \mbox{ } \mbox{ }  \langle \gamma_{\rm{F}} \rangle_{\rm{0}}
  \mbox{ } \mbox{ }$ &
$ \mbox{ } \mbox{ }  \langle \gamma_{\rm{F}}\rangle_{\rm{sdf}}
  \mbox{ } \mbox{ }$ &
$ \mbox{ } \mbox{ }  \langle v^2 \rangle_{\rm{sdf}}
  \mbox{ } \mbox{ }$ &
                                                                                
$ \mbox{ } \mbox{ }  \langle \gamma_{\rm{F}}\rangle_{\rm{0}}
  \mbox{ } \mbox{ }$ &
$ \mbox{ } \mbox{ }  \langle \gamma_{\rm{F}}\rangle_{\rm{sdf}}
  \mbox{ } \mbox{ }$ &
$ \mbox{ } \mbox{ }  \langle v^2 \rangle_{\rm{sdf}}
  \mbox{ } \mbox{ }$ \\
                                                                                
%$ \mbox{ } \mbox{ }  \langle \gamma_{\rm{F}}\rangle_{\rm{0}}
%  \mbox{ } \mbox{ }$ &
%$ \mbox{ } \mbox{ }  \langle \gamma_{\rm{F}}\rangle_{\rm{sdf}}
%  \mbox{ } \mbox{ }$ &
%$ \mbox{ } \mbox{ }  \langle v^2 \rangle_{\rm{sdf}}
%  \mbox{ } \mbox{ }$ \\
\hline\\[-7truept]
                                                                                
$0.08$ & $0.00$ & $0.01$ & $0.67$ &
         $0.00$ & $0.03$ & $0.69$ \\
%         $0.00$ & $0.01$ & $0.66$  \\
                                                                                
$0.2$ &  $0.00$ & $0.01$ & $0.67$ &
         $0.00$ & $0.03$ & $0.68$ \\
%         $0.00$ & $0.01$ & $0.67$  \\
                                                                                
$0.4$ &  $0.00$ & $0.01$ & $0.67$ &
         $0.01$ & $0.03$ & $0.67$ \\
%         $0.00$ & $0.01$ & $0.67$  \\
                                                                                
\hline\\[-9truept]
\hline\\[-8truept]
\end{tabular}
\end{center}
\caption{The average of the friction corrections, $\gamma_{\rm F}$,
calculated by the iterative
self-consistent (0) and space-dependent (sdf)
approaches for the MHO and sinusoidal potentials.
The resulting temperatures are also included for
the space-dependent friction.
In all cases the temperature is 2/3
(in units of a standard temperature, $k_{\rm b} T_0$),
the variance, $\sigma^2=0.22$, and the correlation
time, $\tau_{\rm c}=1$.
}
\label{tb:mho}
\end{table*}

The variance and correlation time
for both potentials is 0.22 and 1, respectively.
The resulting
temperatures, ($k_{\rm B}T\equiv \langle v^2 \rangle$), are also 
listed for the space-dependent approach.  
The friction correction in the self-consistent method
ensures equipartition by definition, and therefore, is not listed.
The magnitude of the SDF
for all values of $\tau_{\rm c}$ follow accordingly;
however this is the only value with respect to the given variance
for which any deviation from equipartition is observed.
As can be seen, 
both the self-consistent and space-dependent components of the total
friction for each potential provide negligible contributions for this
variance since 
the magnitude of the fluctuations
in the barrier height are relatively small.
Therefore the total friction is a sum of 
a large thermal component, and a space-dependent contribution.
The slight differences in the magnitudes of the SDF for the 
two potentials can be attributed to the piecewise nature of the MHO potential.
The particles spend most of the simulation time in the wells which
do not fluctuate.  
A contribution to the total friction from the space-dependent term is 
included only when the energetically-limited particle 
accumulates enough energy to
explore the upper portion of the MHO potential.

To further explore the accuracy of
the space-dependent approach, the sinusoidal
potential has been studied with a ten-fold increase in the variance from
0.22 to 2.2.  
%This corresponds to fluctuations in the barrier heights of
%approximately 25 \% around the average.
The values of the friction correction
from these simulations are listed in Table~\ref{tb:sin}.
%\begin{longtable}{cc}
\begin{table*}[t]
\begin{center}
\begin{tabular}{ccccccccccccc}
%\begin{longtable}{ccccccccccccc}
\hline \\[-9truept]
\hline \\[-8truept]
\multicolumn{1}{c}{} &
\multicolumn{3}{c}{
    \underline{$
                 \mbox{ }\mbox{ }\mbox{ }\mbox{ }\mbox{ }\mbox{ }
                 \mbox{ }\mbox{ }\mbox{ }\mbox{ }\mbox{ }\mbox{ }
                 \tau_{\rm c}=10^{-1}
                 \mbox{ }\mbox{ }\mbox{ }\mbox{ }\mbox{ }\mbox{ }
                 \mbox{ }\mbox{ }\mbox{ }\mbox{ }\mbox{ }\mbox{ }
   $} }&
\multicolumn{3}{c}{
   \underline{$
                 \mbox{ }\mbox{ }\mbox{ }\mbox{ }\mbox{ }\mbox{ }
                 \mbox{ }\mbox{ }\mbox{ }\mbox{ }\mbox{ }\mbox{ }
                 \tau_{\rm c}=10^0
                 \mbox{ }\mbox{ }\mbox{ }\mbox{ }\mbox{ }\mbox{ }
                 \mbox{ }\mbox{ }\mbox{ }\mbox{ }\mbox{ }\mbox{ }
   $} }&
\multicolumn{3}{c}{
   \underline{$
                 \mbox{ }\mbox{ }\mbox{ }\mbox{ }\mbox{ }\mbox{ }
                 \mbox{ }\mbox{ }\mbox{ }\mbox{ }\mbox{ }\mbox{ }
                 \tau_{\rm c}=10^1
                 \mbox{ }\mbox{ }\mbox{ }\mbox{ }\mbox{ }\mbox{ }
                 \mbox{ }\mbox{ }\mbox{ }\mbox{ }\mbox{ }\mbox{ }
   $} }\\
$\gamma_{\rm th}$ &
$ \mbox{ }   \mbox{ }
  \langle \gamma_{\rm{F}} \rangle_{\rm{0}}
  \mbox{ } \mbox{ } $ &
$ \mbox{ }  \mbox{ }
  \langle \gamma_{\rm{F}}\rangle_{\rm{sdf}}
  \mbox{ }  \mbox{ }$ &
$ \mbox{ }   \mbox{ }
  \langle v^2 \rangle_{\rm{sdf}}
  \mbox{ }  \mbox{ }$ &
$ \mbox{ }   \mbox{ }
  \langle \gamma_{\rm{F}}\rangle_{\rm{0}}
  \mbox{ } \mbox{ } $ &
$  \mbox{ } \mbox{ }
  \langle \gamma_{\rm{F}}\rangle_{\rm{sdf}}
  \mbox{ } \mbox{ }$ &
$ \mbox{ }  \mbox{ }
  \langle v^2 \rangle_{\rm{sdf}}
  \mbox{ }  \mbox{ }$ &
$  \mbox{ } \mbox{ }
  \langle \gamma_{\rm{F}}\rangle_{\rm{0}}
  \mbox{ }   \mbox{ }$ &
$   \mbox{ } \mbox{ }
\langle \gamma_{\rm{F}}\rangle_{\rm{sdf}}
  \mbox{ } \mbox{ } $ &
$ \mbox{ }  \mbox{ }
  \langle v^2 \rangle_{\rm{sdf}}
  \mbox{ }   \mbox{ }$ \\
\hline\\[-7truept]
$0.08$ & $0.04$ & $0.28$ & $0.72$ &
         $0.05$ & $0.29$ & $0.74$ &
         $0.01$ & $0.28$ & $0.68$  \\
$0.2$ & $0.04$ & $0.28$ & $0.71$ &
        $0.05$ & $0.29$ & $0.72$ &
        $0.01$ & $0.28$ & $0.68$  \\
$0.4$ & $0.04$ & $0.28$ & $0.70$ &
        $0.06$ & $0.29$ & $0.71$ &
        $0.01$ & $0.28$ & $0.67$  \\
\hline\\[-9truept]
\hline\\[-8truept]
\end{tabular}
%end{longtable}
\end{center}
\caption{The average of the friction corrections, $\gamma_{\rm F}$,
calculated by the iterative self-consistent (0)
and space-dependent approaches (sdf) for the sinusoidal potential.
The resulting temperatures are also included for
the space-dependent friction method.
The temperature is 2/3 in all cases and the variance, $\sigma^2=2.2$.
}
\label{tb:sin}
\end{table*}
%\end{longtable}

The displayed correlation times, $\tau_{\rm c}$, are those that
exhibit the largest resonant activation.  
Consequently, if memory effects in the barrier heights are 
important in determining the friction constant, 
it should be manifested here.  
Although not shown for brevity, 
outside this region of the correlation time,
the magnitude of the deviations from equipartition 
%for the space-dependent friction 
decrease rapidly, but the size of the space-dependent components
remains roughly constant.
Similarly, the corresponding corrections arising in the
self-consistent method also approach zero.
As can be seen from Table~\ref{tb:sin}, the space-dependent approach 
results in a correction that is roughly constant for all values of 
the correlation time, while the iterative approach does exhibit some 
variation with $\tau_{\rm c}$.  
This is the expected result since the space-dependent
friction assumes the fluctuations in the potential are local
and therefore, ignores any correlation in
the barrier heights. 
The iterative approach, however, is capable of
incorporating the memory of the potential into the friction correction, 
but only in an average manner.  
As a consequence, significant deviations from equipartition may be 
observed when simulations are performed with a 
space-dependent friction that ignores the correlation effects, as
illustrated by this extreme example.

\begin{figure}[ht]
\begin{center}
\includegraphics*[width=7.5cm]{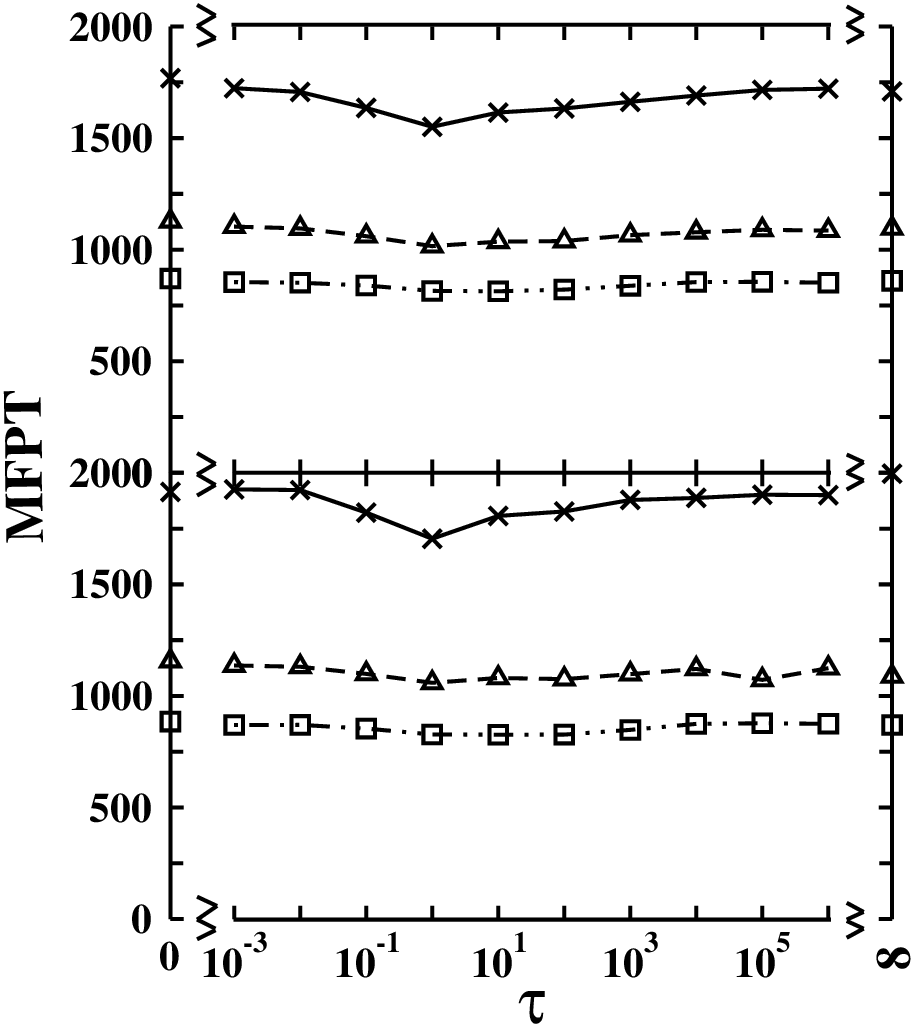}
\end{center}
\caption{The mean first-passage times (MFPT)
for a particle diffusing across the MHO stochastic potential
are displayed for two possible scenarios of the
dissipative mechanism.
The top panel uses space-dependent friction, and the bottom
displays the uniform friction determined by the self-consistent method.
The variance for both is 0.05, and
the three lines correspond to values of the thermal
friction of 0.08 (solid curve with x symbols), 0.2 (dashed curve
with triangles), and 0.4 (dot-dashed curve with squares).
The symbols on the broken axis represent the
numerically calculated MFPTs at the limits of the correlation time.
}
\label{fig:mfpt0.05}
\end{figure}

\begin{figure}[ht]
\begin{center}
\includegraphics*[width=7.5cm]{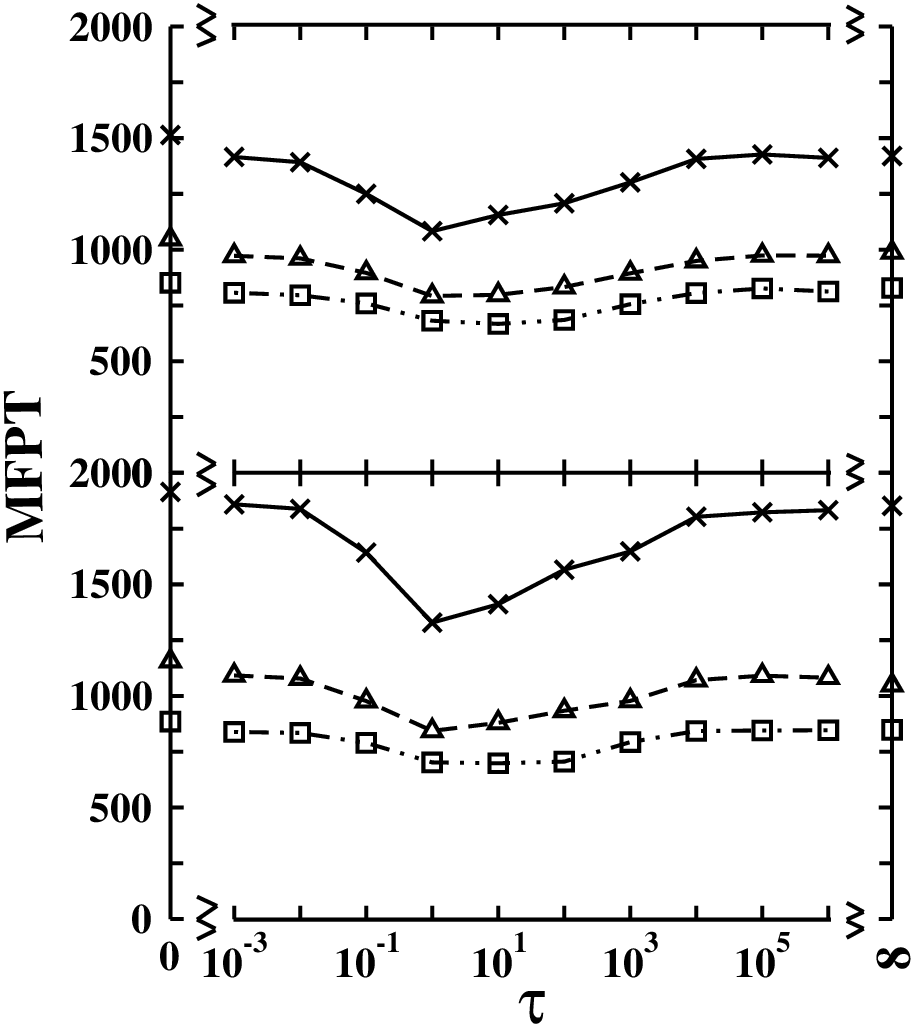}
\end{center}
\caption{The mean first-passage times (MFPT)
for a particle diffusing across the MHO stochastic potential
are displayed for two possible scenarios of the
dissipative mechanism.
The parameters are the same as in Fig.~\ref{fig:mfpt0.05}, except the
variance is 0.22.
}
\label{fig:mfpt0.22}
\end{figure}

Figs.~\ref{fig:mfpt0.05} and ~\ref{fig:mfpt0.22} display the 
MFPTs obtained for the MHO potential 
with the results from the space-dependent and self-consistent 
approaches in the top and bottom panels, respectively. 

%%%%%%%%%%%%%%%%%%%%%%%
%\newpage
                                                                                
%%%%%%%%%%%%%%%%%%%%%%%
%%%%%%%%%%%%%%%%%%%%%%%
%\newpage
 
\begin{figure}[t]
\begin{center}
\includegraphics*[width=7.5cm]{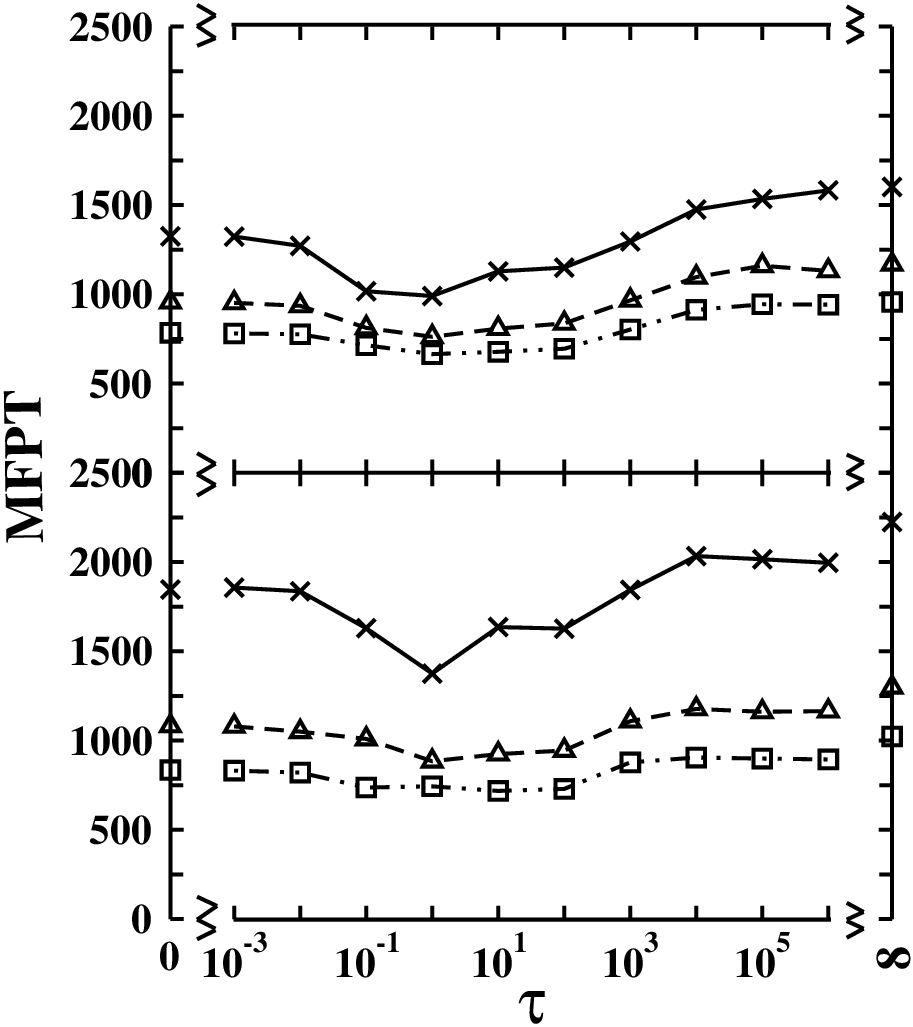}
\end{center}
\caption{The mean first-passage times (MFPT)
for a particle diffusing across the sinusoidal stochastic potential
are displayed for two possible scenarios of the
dissipative mechanism.
Other than for the change from the MHO to the sinusoidal potential,
the parameters are the same as in Fig.~\ref{fig:mfpt0.22}.
}
\label{fig:mfptsin}
\end{figure}

The results in Fig.~\ref{fig:mfpt0.05} have been calculated using a variance
of $\sigma^2=0.05$, while those in Fig.~\ref{fig:mfpt0.22} use
$\sigma^2=0.22$.
The corresponding results for the sinusoidal potential using 
a variance of 0.22 can be seen in Fig.~\ref{fig:mfptsin}.
%%%%%%%%%%%%%%%%%%%%%%%
%%%%%%%%%%%%%%%%%%%%%%%
The values on the broken axis represent the numerically calculated
MFPTs in the limits of correlation time, $\tau_{\rm c}$.
In the zero-correlation time limit, 
the fluctuations in the potential are so rapid that the 
particle effectively experiences the average, stationary potential, 
from which the dynamics were calculated.  
In the limit of infinite correlations, 
fluctuations in the potential are nonexistent, 
and therefore the particle experiences a single realization 
of the potential with constant barrier heights determined by the 
initial value sampled from the distribution. 
The MFPTs displayed in Fig.~\ref{fig:mfpt0.22}
obtained with a larger variance alters
the magnitude of the resonant activation, but 
influences the results for the two approaches equally.
The results from the simulations with a space-dependent friction 
are systematically shifted to lower MFPTs as seen in all three figures.  
This trend is most readily explained through by the 
trends in Table.~\ref{tb:mho}.  
In the low friction regime, an increase in the friction
increases the corresponding rate of transport.
The average space-dependent contribution is always larger than
its respective mean field counterpart, 
and is expected to have the largest effect
on the results with the smallest thermal friction.
The fluctuations present along the entire reaction coordinate
of the sinusoidal potential do not appear to have a dramatic effect
on the dynamics.  
The results in Fig.~\ref{fig:mfptsin} for the sinusoidal potential follow
the same trend as those in Figs.~\ref{fig:mfpt0.05} and \ref{fig:mfpt0.22}
for the MHO potential indicating that the SDF approach is capable of 
adequately describing the fluctuations in the system.
Aside from the shift, the general behavior of the MFPT is adequately
reproduced by both methods, particularly at larger values
of the thermal friction when the space-dependent component becomes
less significant.  
At this level of description,
each of the two approaches for constructing the friction
are capable of capturing the essential dynamics of the system.
However, 
some advantage is gained by using the self-consistent
method because it ensures the system is kept at constant temperature 
for all values of the correlation time throughout the simulation,
while the space-dependent approach may lead to deviations in extreme cases.
The most significant difference between the two methods can be
seen at intermediate correlation times, in which the resonant
activation observed from the iterative approach is slightly 
more pronounced. 
This can particularly be seen in the MFPTs
when the friction case takes on the smallest value of $\gamma_{\rm{th}}=0.08$.
Since the resonant activation arises from correlations in the barrier heights, 
it is not surprising that simulations incorporating a friction
capable of accounting for this phenomenon can have a noticeable
impact on the dynamics, 
even if it does so only in an average manner.

\section{Conclusions}\label{sec:conclusion}

The space-dependent friction arising from the 
presence of a secondary (external) stochastic potential in 
the Langevin equation has been explicitly derived for two simple classes of
the stochastic potentials.
The numerical results are
in excellent agreement with analytic expressions 
describing the space-dependent friction. 
The resulting dynamics have been compared to those obtained using
an alternate approach in which a uniform correction is calculated 
self-consistently.
Although the latter approach does effectively include the 
time correlation between the barrier fluctuations at long times,
the former does not in any sense.
This neglect may result in deviations from equipartition in 
some extreme cases.  
However, both approaches are capable of capturing
the essential dynamics of the system and lead
to the now-expected resonant activation phenomenon.
Consequently, the central result of this paper is that the Langevin
dynamics of a particle under external stochastic potentials
can be properly dissipated by a single uniform renormalized friction
without loss of qualitative (and often quantitative) accuracy.

The role of the memory time in
an external stochastic potential acting on a particle 
described by a generalized Langevin equation of motion
is still an open question.
In this limit, there would presumably be an interplay between the
memory time of the thermal friction and that of the stochastic potential.
When the latter is small compared to the former, the quasi-equilibrium
condition central to this work would no longer be satisfied by the particle,
and hence it is expected that a non-uniform (and time-dependent) friction
correction would then be needed.

\section{Acknowledgments}%

RH gratefully acknowledges Abraham Nitzan for
an insightful question whose answer became this paper.
This work has been partially supported by a National Science Foundation Grant,
No.~NSF 02-123320.  
The Center for Computational Science and Technology
is supported through a Shared University Research (SUR)
grant from IBM and Georgia Tech.
Additionally, RH is the Goizueta Foundation Junior Professor.

\onecolumngrid
\section{Appendix}%

The piecewise nature of the MHO potential results in a
piecewise form for the associated SDF.
Although incoherent, every barrier gives rise to the same averages,
and hence the procedure needs to be carried out only over a 
small region defined by the closed interval, $[x_m^0,x_m^\ddagger]$.  
The limits of integration over this region can be 
determined from the expression for the connection points
\begin{equation}
x_m^-=-\frac{k_0\lambda}{2k_0-2k_m^\ddagger}+m\lambda
\;,
\end{equation}
where $k_m^{\ddagger}=-(k_0+\eta(t))$. 
This can equivalently be expressed as
\begin{equation}
\eta(t)=-\frac{k_0\lambda}{2(x_m^--m\lambda)}-2k_0
\;.
\end{equation}
At the top of the barrier, when $x_m^-=x_m^\ddagger,~
\mbox{$\eta(t)=\infty$}.$
In the intermediate region for arbitrary $x$,
\begin{eqnarray}
\eta(t) & = & -\frac{k_0\lambda}{2(x-m\lambda)}-2k_0 \nonumber \\
        & \equiv & \eta^*
\;.
\end{eqnarray}
Otherwise, at the minimum when $x_m^-=x_m^0,~\mbox{$\eta(t)=-k_0$} $.

Although it is apparent from the expression for the barrier height
that the corresponding distribution is non-Gaussian, the 
resulting forces are Gaussian with the probability given by Eq.~\ref{eq:prob}. 
The average force for a given $x$ is simply the weighted average of 
the forces when $x$ is in the respective regions,
$(x_m^0,x_m^-)$ and $(x_m^-,x_m^{\ddagger})$, 
which correspond to $\eta$ regions of
$(-k_0,\eta^*)$ and $(\eta^*,\infty)$. 
The resulting integral for the average value of $F(x;t)$ is now:
\begin{equation}
\langle F_{\rm U}(x;t) \rangle= \frac{\int_{-k_0}^{\eta^*}\!d\eta\, F(x)P(\eta)
                                +\int_{\eta^*}^{\infty}\!d\eta\,F(x)P(\eta)}
                                {\int_{-k_0}^{\infty}\!d\eta\,P(\eta)}
\;.
\end{equation}
Here, one must be careful 
in determining which portion of the force to use in the above
equation.  
For example, when $\eta<\eta^*$, 
the majority of the force is due to the barrier portion of the potential, 
not the well component.  
The average can thus be expressed as
\begin{equation}
\langle F_{\rm U}(x;t) \rangle= -\int_{\eta^*}^{\infty}\!d\eta\, 
                               k_0(x-x_m^0)P^\prime(\eta)
                               +\int_{-k_0}^{\eta^*}\!d\eta\,
                               (k_0+\eta)(x-x_m^{\ddagger})P^\prime(\eta)
\;,
\end{equation}
where $P^\prime(\eta)$ is defined through the 
normalization condition, $\it{i.e.}$,
\begin{equation}
\int_{-k_0}^{\eta^*}\!d\eta\,P^\prime(\eta)+
\int_{\eta^*}^{\infty}\!d\eta\,P^\prime(\eta)\equiv 1
\;,
\end{equation}
which leads to the probability distribution
\begin{equation}
P^\prime(\eta)=\frac{2}{\sqrt{2\pi\sigma^2}}
               \frac{\exp\left({-\frac{\eta^2}{2\sigma^2}}\right)}
               {1+\erf\left(\frac{k_0}{\sqrt{2\sigma^2}}\right)}
\;,
\end{equation}
where $\erf(x)$ is the standard error function.
Use of the normalization condition reduces the
average force to
\begin{equation}
\langle F_{\rm U}(x;t) \rangle  =  k_0(x-x_m^\ddagger)-
                             [k_0(x-x_m^0)+k_0(x-x_m^{\ddagger})]
                             \int_{\eta^*}^{\infty}\!d\eta\,
                             P^\prime(\eta)  \nonumber 
                             \mbox{} + (x-x_m^{\ddagger})\int_{-k_0}^{\eta^*}
                             \!d\eta\, \eta P^\prime(\eta)
\;.
\end{equation}
The remaining integrals are readily computed; the explicit form of 
the average force is
\begin{eqnarray}
\langle F_{\rm U}(x;t) \rangle & = & k_0(x-x_m^\ddagger)-
                                   \left( k_0 (2x-x_m^0-x_m^\ddagger)\right)
                                   \left(\frac{1-\erf\left(\frac{\eta^*}
                                   {\sqrt{2\sigma^2}}\right)}{1
                                    +\erf\left(\frac{k_0}{\sqrt{2\sigma^2}}
                                    \right)}\right) \nonumber \\
                               &   & \mbox{}+(x-x_m^\ddagger)
                                     \sqrt{\frac{2\sigma^2}{\pi}}
                                     \left[\frac{\exp\left(-
                                     \frac{k_0^2}{2\sigma^2}\right)-
                                     \exp\left(-
                                     \frac{(\eta^*)^2}{2\sigma^2}\right)}
                                     {1+\erf\left(
                                    \frac{k_0}{\sqrt{2\sigma^2}}\right)}\right]\;.               
\end{eqnarray}

The second quantity to be computed is the average of the 
square of the force, and the derivation follows that (above) of
the average force.
The limits of integration are the same 
and the Gaussian integrals can be calculated in the same manner.
Again using the normalization requirement, 
the first integral is eliminated such that
\begin{eqnarray}
\langle F_{\rm U}(x;t)^2 \rangle & = & k_0^2(x-x_m^\ddagger)^2 +
                                       [k_0^2(x-x_m^0)^2-
                               k_0^2(x-x_m^\ddagger)^2]
                               \int_{\eta^*}^{\infty}\!d\eta\, 
                               P^\prime(\eta)
                               \nonumber   \\
                         &   & \mbox{}+2k_0(x-x_m^\ddagger)^2
                               \int_{-k_0}^{\eta^*}
                               \!d\eta\,\eta P^\prime(\eta)+(x-x_m^\ddagger)^2
                               \int_{-k_0}^{\eta^*}\!d\eta\,\eta^2 
                               P^\prime(\eta)   
\;.
\end{eqnarray}
The first two integrals are the same as before, 
and the third can be obtained with little effort.  
The resulting mean squared force is
\begin{eqnarray}
\langle F_{\rm U}(x;t)^2 \rangle & = & k_0^2(x-x_m^\ddagger)^2+\left(k_0^2
                               (x-x_m^0)^2-k_0^2(x-x_m^\ddagger)^2\right)
                               \left(\frac{1-\erf\left(
                               \frac{\eta^*}{\sqrt{2\sigma^2}}
                               \right)}{1+\erf\left(\frac{k_0}
                               {\sqrt{2\sigma^2}}\right)}\right)
                               \nonumber   \\
                         &   & \mbox{} +\frac{4k_0\sigma^2}{\sqrt{2
                               \pi\sigma^2}}(x-x_m^{\ddagger})^2
                               \left[\frac{ \exp\left(-
                               \frac{k_0^2}{2\sigma^2}\right)-
                               \exp\left(-\frac{(\eta^*)^2}{2\sigma^2}\right)}
                               {\left(1+\erf\left(\frac{k_0}
                               {\sqrt{2\sigma^2}}\right)\right)}\right]
                               \nonumber   \\
                         &   & \mbox{} +\sigma^2(x-x_m^{\ddagger})^2
                               \left[ \frac{
                               \erf\left(\frac{\eta^*}{\sqrt{2
                               \sigma^2}}\right)
                               +\erf\left( \frac{k_0}{\sqrt{2
                               \sigma^2}} \right)}
                               {1+\erf\left(\frac{k_0}{\sqrt{2
                               \sigma^2}}\right)}\right]\nonumber\\
                         &   & \mbox{} -\sqrt{\frac{2
                               \sigma^2}{\pi}}(x-x_m^\ddagger)^2
                               \left[ \frac{\eta^*\exp\left(-
                               \frac{(\eta^*)^2}{2\sigma^2}\right) 
                               +k_0\exp\left(-\frac{k_0^2}{2\sigma^2}\right)}
                               {1+\erf\left(\frac{k_0}{2
                               \sigma^2}\right)}\right]
\;.
\end{eqnarray}
The SDF for the MHO potential is then obtained by 
appropriate substitutions into Eq.~\ref{eq:FDT}.

\twocolumngrid
\bibliography{j,surf,tst,miller2,hern,voth,flucbar,mfpt,liquid,sdf}

\end{document}